\documentclass[11.5pt]{article}

\usepackage[english]{babel}

\usepackage[a4paper,top=2cm,bottom=2cm,left=3cm,right=3cm,marginparwidth=1.75cm]{geometry}

\usepackage[backend=biber,style=nature,doi=true]{biblatex}
\usepackage{amsmath}
\usepackage{amssymb}
\usepackage{graphicx}
\usepackage[colorlinks=true, allcolors=blue]{hyperref}
\usepackage[auth-sc, affil-it]{authblk}
\usepackage{changepage}

\usepackage{subcaption}
\usepackage{csquotes}
\usepackage{acro}
\usepackage{siunitx}

\usepackage{booktabs}
\usepackage{colortbl}
\usepackage{xcolor}
\usepackage{xfrac}

\usepackage{caption}
\newcommand{\auth}[1]{\textit{#1}}
\newcommand{\auths}[1]{\textit{#1 et al.}}

\usepackage{parskip}
\DeclareAcronym{SIFs}{
  short=SIFs,
  long=SIFs,
}

\DeclareAcronym{DIC}{
  short=DIC,
  long=digital image correlation,
}

\DeclareAcronym{FEA}{
  short=FEA,
  long=finite element analysis,
}

\DeclareAcronym{ODM}{
  short=ODM,
  long=over-deterministic method,
}
\DeclareAcronym{SECT}{
  short=SECT,
  long=single-edge cracked tension,
}

\DeclareAcronym{CT}{
  short=CT,
  long=compact tension,
}

\providecommand{\keywords}[1]
{
  \small	
  \textbf{\textit{Keywords---}} #1
}

\title{Advanced crack tip stress analysis using interaction integrals in high-resolution digital image correlation fields}
\author[1,*]{Florian Paysan}
\author[1]{David Melching}
\author[1]{Eric Breitbarth}
\affil[1]{German Aerospace Center (DLR), Institute of Materials Research, Linder Hoehe, 51147 Cologne, Germany.}
\affil[*]{Corresponding author: Florian.Paysan@dlr.de}

\date{\today}

\begin{document}
\maketitle

%%%%%%%%%%%%%%%%%%%%%%%%%%%%%%%%%%%%%%%%%%%%%%%%%%%%%%%%%%%%%%%%%%%%%%%%%%%%%%%%%%%
% TO DO
% - fehlende Referenzen ergänzen und Referenzen harmonisieren.
%%%%%%%%%%%%%%%%%%%%%%%%%%%%%%%%%%%%%%%%%%%%%%%%%%%%%%%%%%%%%%%%%%%%%%%%%%%%%%%%%%%

\begin{abstract}
The link between microscopic mechanisms and macroscopic behaviour, represented by the $da/dN-\Delta K$ curve, plays an increasingly important role in relating the fatigue crack growth curve required for component design to the underlying physics. High-resolution digital image correlation (HR-DIC) allows for in-depth analysis of microscopic fatigue crack growth mechanisms, but is rarely used to determine the SIF of the crack tip. This paper examines the applicability of the interaction integral in HR-DIC data and identifies factors that should be considered when evaluating the integral results.

A major influence is the integration near the PZ, which leads to an erroneous increase in the calculated SIF result. In addition, the large integration path gaps required in HR-DIC around the crack path significantly hinder accurate results. The effect of the crack face contact is overall small. While it slightly increases the SIF result, it does not correlate with the crack opening load $K_\mathrm{op}$.
\end{abstract}


%%%%%%%%%%%%%%%%%%%%%%%%%%%%%%%%%%%%%%%%%%%%%%%%%%%%%%%%%%%%%%%%%%%%%%%%%%%%%%
\section{Introduction} \label{sec:intro}
%%%%%%%%%%%%%%%%%%%%%%%%%%%%%%%%%%%%%%%%%%%%%%%%%%%%%%%%%%%%%%%%%%%%%%%%%%%%%%

% general importance of the topic
The primary cause of failure in aircraft components is the formation of fatigue cracks due to non-constant multi-axial loading \cite{Schijve.2009}. In materials science, fatigue crack growth is driven by the interaction of damaging and protective mechanisms near the crack tip \cite{Ritchie.1999}. Despite extensive research, mechanisms like crack closure, branching, or deflection, and their effects on crack propagation, such as crack tip stress, remain poorly understood \cite{Jones.2014}. The challenge lies in experimentally identifying these mechanisms, as they operate at the microscopic level and are often undetectable through post-mortem fracture surface analysis. Thus, high-resolution time-series imaging of the crack tip region is essential throughout the fatigue crack growth process \cite{Chowdhury.2016}.

Special focus is given to the examination of the plastic zone (PZ), including both primary and cyclic PZ, as it is not only the site of damaging crack propagation mechanisms, but also because its dimensions and shape reflect the prevailing stress state at the crack tip. Due to experimental limitations, the PZ has been subject of various 2D and 3D finite element simulations \cite{Camas.2017, Camas.2012, Camas.2011, GarciaManrique.2013, GonzalezHerrera.2008}, which found that the shape of the PZ deviates from the dog-bone model, initially proposed by Dugdale \cite{Dugdale.1959}. \auths{Besel} \cite{Besel.2016} observed that the shape of the PZ underwent changes at both the free surface and within the volume of the specimen. They attributed these observations to the interaction between the plane strain within the volume and the plane stress at the surface of the specimen. Both conditions depend strongly on the specimen thickness and the maximum crack tip stress \cite{Caputo.2013, Gadallah.2021}.
In recent years, the advent of high-resolution digital image correlation (HR-DIC) has notably increased the number of experimental studies examining the PZ. In 2013, HR-DIC research was conducted by \auths{Carroll} \cite{Carroll.2013}. They analyse the dislocation structures in the plasticised region in proximity to the crack tip. It was observed that the accumulation of local strains within both the PZ and the plastic wake is highly inhomogeneous. Other studies used HR-DIC to examine the wing shape of the PZ. 
\auths{Zhang} \cite{Zhang.2019} examined the effects of overload on the PZ and found that the subsequent deformation zone was significantly smaller. A comparable investigation of the PZ following an overload was conducted by \auths{Vasco-Olmo} \cite{VascoOlmo.2018}. The results agreed with those of \auths{Zhang} \cite{Zhang.2019}, indicating that the temporary delay in crack growth rate was significantly associated with the overload-induced PZ. \auths{Besel} \cite{Besel.2016} concentrated on examining the shape of the PZ in relation to varying sheet thicknesses and crack tip stresses. Their findings elucidated a clear correlation between the PZ geometry and the sheet thickness supporting their numerical studies.

% crack tip stress determination based on dic
Digital image correlation (DIC) has become a crucial tool in fracture mechanics, as it allows to determine crack tip stresses based on measured full-field displacement data near the crack tip, thereby providing valuable insights into the mechanics of fracture. The stress is usually derived from the measured strains by a known constitutive law (e.g. Hooke's law). The currently applied methods to calculate stress tip field descriptors like SIFs or T-stress can be grouped into two classes - invariant integrals and over-deterministic methods. Both require an accurate determination of the crack tip position. While \auth{Gehri} \cite{Gehri.2020} uses the displacements of the two crack flanks as the basis for evaluation, \auth{Rethore} \cite{Rethore.2015} derived an iterative correction formula to refine the crack tip position of a mode I crack based on an initial estimate and the over-deterministic method. \auths{Melching} \cite{Melching.2024} used physical deep symbolic regression to find more general iterative correction formulas for mode II and mode-I-dominated  mixed mode loading scenarios. Furthermore, \auths{Strohmann} \cite{Strohmann.2021} employ a convolutional neural network for crack tip detection in noisy DIC data and \auths{Melching} \cite{Melching.2022} improved upon this by using explainable machine learning to select a model which focuses neural attention on the crack tip field.

For the purpose of calculating crack tip stresses from DIC data, the $J$-integral has become particularly well-established \cite{Gonzales.2017, Breitbarth.2019, Molteno.2015}. Given the complex cyclic behaviour of the material within the PZ, all studies have in common that integration paths are placed outside the PZ, justifying the use of Hooke's law as the material model for the $J$-integral formulation. \auths{Gonzales} \cite{Gonzales.2017} calculate SIFs during an overload for different integration path geometries. Although their integration paths contain gaps to exclude the erroneous singular strain surrounding the crack path, their $K_\mathrm{I}$ results converge. \auths{Breitbarth} \cite{Breitbarth.2019} recommend using multiple integration paths to negate the impact of DIC noise. Additionally, they assert that the $J$-integral does not require an exact crack tip position. \auths{Molteno} \cite{Molteno.2015} introduces a new decomposition method to extract mixed-mode SIFs from the classical $J$-integral calculation. He investigates different integration paths with radii ranging from $r=3\, \mathrm{mm}$ to $r=12 \, \mathrm{mm}$. He found that the $K_\mathrm{I-III}$-results start to converge from $r>4\, \mathrm{mm}$. The interaction integral itself belongs to the methods of linear-elastic fracture mechanics and has been succesfully employed to DIC data to calculate mixed-mode SIFs \cite{Breitbarth.2019, Rethore.2005}. In his studies, \auths{Breitbarth} \cite{Breitbarth.2019} conclude that DIC noise has a more significant role in determining the final results than inaccuracies regarding the position of the crack tip. \auths{Rethore} \cite{Rethore.2005} propose that the interaction integral should be evaluated as a domain integral, which has the advantage of reducing sensitivity to errors made by the DIC displacement calculation. 

However, in all studies mentioned above macroscopic DIC data is used, which does not permit an in-depth analysis of local fatigue crack growth mechanisms as is possible with HR-DIC data. Due to the large distance between the crack tip and the integration contours or paths, respectively, effects such as plasticity of the PZ and plastic wake, as well as crack closure contact, are neglected in macroscopic DIC. 

\begin{figure}[ht]
	\centering
	\includegraphics[width=0.95\textwidth]{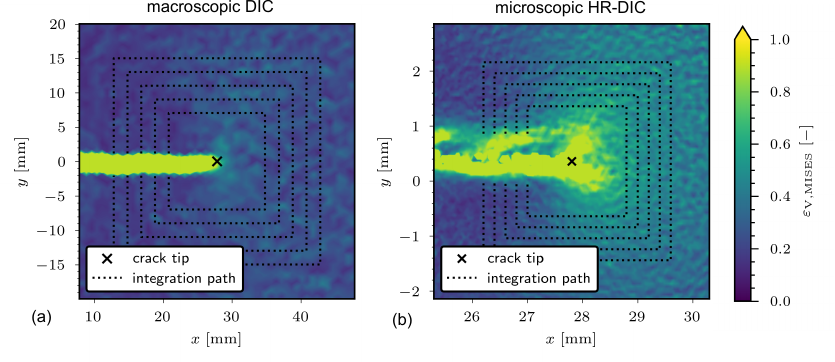}
	\caption{Comparison of (macroscopic) DIC and HR-DIC strain fields showing the equivalent MISES strain: (a) a macroscopic DIC strain field obtained from a 3D measurement and (b) an HR-DIC strain field captured using an optical microscope; both evaluated at $K_\mathrm{norm}=14.9\, \mathrm{MPa\sqrt{m}}$ (maximum load) and $a=27.8\, \mathrm{mm}$; $K_\mathrm{norm}$ is determined by ASTM E647-15 \cite{ASTM.E647} (see equation \ref{eq:mt_LEBM})}
	\label{fig:dic_hrdic_comparison}
\end{figure}

The objective of this paper is to investigate the impact of near-crack-tip effects on SIF results calculated by the interaction integral method using a line integral formulation. To achieve this, this paper investigates the following characteristics which are typical for applying line integrals in HR-DIC displacement field data: 
\begin{itemize}
    \item Given the limited measurement range of HR-DIC systems, this study examines the impact of the near-PZ on the outcome of SIF calculations at the crack tip.
    \item The integration path gap in HR-DIC evaluations is considerably larger than the overall integration path length. This study looks at how the gap affects the SIF result and the impact of integrating parts of the path within the plastic wake.
    \item Finally, it is examined how the SIF result of the interaction integral reflects the crack face contact due to crack closure.
\end{itemize}  
In order to distinguish between the individual influences, a finite element simulation is employed to model effects such as the elastic-plastic deformation of the PZ and the plastic wake, as well as crack closure. The observed characteristics within the result of the interaction integral are then compared with those derived from the HR-DIC data. An MT-160 specimen of 2-mm-thick aluminum alloy AA2024-T3, widely used in the aircraft industry, is utilized in this study. To illustrate the data sets, Figure \ref{fig:dic_hrdic_comparison} presents a comparison between two different strain fields: a macroscopic DIC strain field, commonly used to determine SIFs through line integrals, and a high-resolution DIC (HR-DIC) strain field.

%%%%%%%%%%%%%%%%%%%%%%%%%%%%%%%%%%%%%%%%%%%%%%%%%%%%%%%%%%%%%%%%%%%%%%%%%%%%%%
\section{Methodology} \label{sec:method}
%%%%%%%%%%%%%%%%%%%%%%%%%%%%%%%%%%%%%%%%%%%%%%%%%%%%%%%%%%%%%%%%%%%%%%%%%%%%%%

\subsection{Interaction integral}
\label{sec:interaction_integral}
In order to distinguish between mode-I and mode-II crack loading conditions, \auths{Stern} \cite{Stern.1976} and \auths{Yau} \cite{Yau.1980} developed the so-called interaction integral $J^{(a,b)}$, based on the classical $J$-integral implementation. The mathematical definition of the interaction integral is given in Formula \ref{eq:interaction_integral}. The integral evaluates the stresses $\sigma$, strains $\varepsilon$, and displacements $u$ along an integration path $\Gamma_r$. $n$ defines the outward unit normal to the integration path $\Gamma_r$. In general, the interaction integral is based on the superposition of two loading cases $^{(1)}$ and $^{(2)}$, where in practice $(1)$ is the actual measured field and $(2)$ is an arbitrary auxiliary field.
\begin{equation}
	\label{eq:interaction_integral}
	J^{(1,2)} = \lim _{r \rightarrow 0} \int_{\Gamma_r} \left[
	\sigma_{m n}^{(2)} \varepsilon_{m n}^{(1)} \delta_{1 j} - \left(\sigma_{i j}^{(1)} u_{i, 1}^{(2)} - \sigma_{i j}^{(2)} u_{i, 1}^{(1)}\right) \right] n_j \, \mathrm{d}s
\end{equation}
More precisely, two different auxiliary fields $^{(2a)}$ and $^{(2b)}$ are defined. The resulting system of equations allows to determine $K_\mathrm{I}^{(1)}$ and $K_\mathrm{II}^{(1)}$:
\begin{flalign}
	\begin{aligned}
		& K_{\mathrm{I}}^{(1)}=\frac{E}{K^2} \cdot\left[K_{\mathrm{II}}^{(2\mathrm{a})} \cdot J^{(1,2\mathrm{b})}-K_{\mathrm{II}}^{(2\mathrm{b})} \cdot J^{(1,2\mathrm{a})}\right] \\
		& K_{\mathrm{II}}^{(1)}=\frac{E}{K^2} \cdot\left[K_{\mathrm{I}}^{(2\mathrm{b})} \cdot J^{(1,2\mathrm{a})}-K_{\mathrm{I}}^{(2\mathrm{a})} \cdot J^{(1,2\mathrm{b})}\right] \\
		& K^2=K_{\mathrm{I}}^{(2\mathrm{b})} \cdot K_{\mathrm{II}}^{(2\mathrm{a})}-K_{\mathrm{I}}^{(2\mathrm{a})} \cdot K_{\mathrm{II}}^{(2\mathrm{b})}
	\end{aligned}
\end{flalign}
If a pure mode-I loading is used for the auxiliary field $^{(2a)}$ or a mode-II crack loading is used for the auxiliary field $^{(2b)}$, the above equations simplify as follows:
\begin{flalign}
	\begin{aligned}
		& K_{\mathrm{I}}^{(1)}=\frac{E}{K_{\mathrm{I}}^{(2 \mathrm{a})}} \cdot J^{(1,2 \mathrm{a})} \\
		& K_{\mathrm{II}}^{(1)}=\frac{E}{K_{\mathrm{II}}^{(2 \mathrm{b})}} \cdot J^{(1,2 \mathrm{b})}
	\end{aligned}
\end{flalign}

\begin{figure}[ht]
	\centering
	\includegraphics[width=0.3\textwidth]{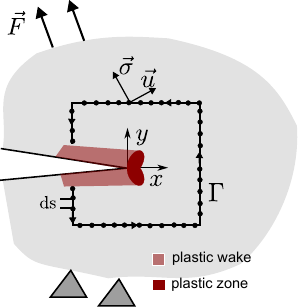}
	\caption{Integration path definition for evaluating the interaction integral in displacement field data; the definition is equivalent to the implementation in \textit{CrackPy} \cite{CrackPy.2022}}
	\label{fig:integral_path}
\end{figure}

DIC measurements only provide the displacement and total strain field as its derivative of the free surface of the MT-160 specimen. In order to calculate the stresses $\sigma$ in Equation \ref{eq:interaction_integral}, a constitutive equation that connects stresses to strains is needed. The local cyclic stress-strain behaviour within the PZ is complex. An incorrect representation of stress-strain behaviour may result in the loss of path independence of the interaction integral as has been demonstrated in case of the $J$-integral \cite{Xu.2022}. Consequently, the integration path is typically positioned outside the PZ within the linear-elastic crack tip field (see Figure \ref{fig:integral_path}), thus justifying the usage of Hooke's linear-elastic constitutive law. Unfortunately, plastic wake, inaccurate strain calculations and scatter around the crack path hinder to close the integration paths. While this gap can be neglected for most cases in macroscopic DIC \cite{Gonzales.2017}, its influence for HR-DIC data needs to be considered.

In this paper, the authors follow the methodology from \auths{Breitbarth} \cite{Breitbarth.2019} to evaluate the interaction integral with DIC displacement field data. The algorithm is implemented in the open-source software package \textit{CrackPy} (\url{https://github.com/dlr-wf/crackpy}), a crack analysis tool written in Python, which offers several methods for crack tip stress field evaluation. The basis for the interaction integral is a rectangular integration path that is discretised over integration points as schematically illustrated in Figure \ref{fig:integral_path}. At each integration point, the strains $\varepsilon$ and stress $\sigma$ are interpolated from the surrounding domain. In most cases, FE or DIC full-field displacement data is provided in the form of unstructured grid data. Consequently, the point information is typically obtained through linear interpolation from the nearest grid points. In this study, the distance between the interpolation points is set to $\SI{0.05}{mm}$, which is identical to the refined element size near the crack tip employed in the FE simulation (see Section \ref{sec:fe_simulation}). Finally, the interaction integral in Equation \ref{eq:interaction_integral} is calculated using numerical trapeze integration method. In the context of DIC data, it is recommended to use multiple integration paths to mitigate the impact of inherent data noise on the integral results.

\subsection{Finite element model}
\label{sec:fe_simulation}
\begin{figure}[ht]
	\centering
	\includegraphics[width=0.95\textwidth]{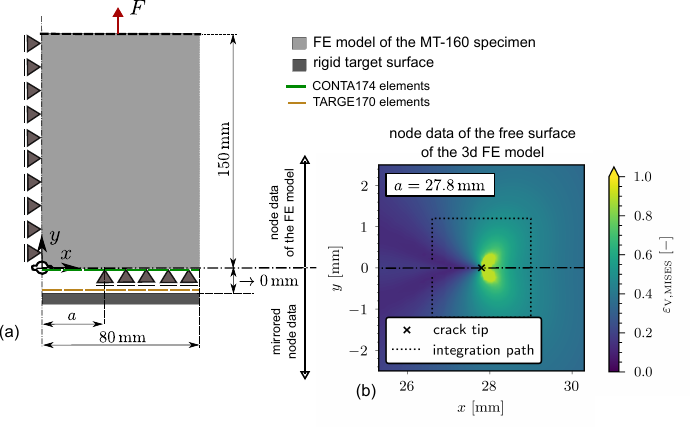}
	\caption{Strategy to evaluate the interaction integral on numerical displacement field data: (a) 1/8 FE model of a MT-160 specimen including its constraints and contact element definition, (b) the displacement node data, partially mirrored, of the free specimen surface}
	\label{fig:fe_model}
\end{figure}

All simulations are conducted using ANSYS Mechanical APDL on a RedHat Linux workstation with dual Intel Xeon Gold 6240 18C CPUs and 256 GB DDR4-2933 RAM. The numerical simulations are based on the FE model from \cite{Paysan.2022}. For completeness, a brief summary of the FE model is provided below:

The FE model is derived from the MT-160 specimen geometry and its general dimensions are displayed in Figure \ref{fig:fe_model}a. To optimize computational efficiency, a 1/8 symmetry model is used, excluding the clamping region and holes to enable mapped meshing. Figure \ref{fig:fe_model}a details the constraint setup. The load $F$ is applied through a pilot node coupled to all top surface nodes. Symmetry planes restrict node movement perpendicular to their planes, with a crack of length $a$ defined at the model's bottom. Symmetry constraints in the interval $x \in [0, a]$ are removed to allow crack surface deformation. 

\begin{figure}[ht]
	\centering
	\includegraphics[width=0.95\textwidth]{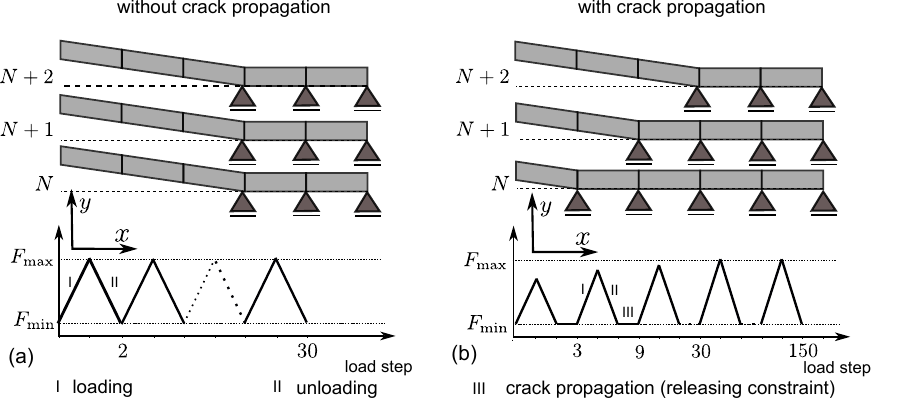}
	\caption{Summary of the current state of research towards the identification of crack closure using compliance methods: (a) measurement methods, (b) crack opening curve related to the corresponding measurement method.}
	\label{fig:crack_growth_algorithm}
\end{figure}

The model uses a mapped mesh strategy with 8-node linear hexahedron SOLID185 elements. The structured mesh is divided into three sections: a global mesh, a refined region near $a$ for detailed analysis, and a transition section connecting both. The dimensions of the refined region depends on the initial size estimate according to Irwin (see \cite{Paysan.2022}) The refined region has elements with an aspect ratio of 1 and an element size of $\SI{0.05}{mm}$ in all three Cartesian directions. Outside the refined section, a flexible element size of $\SI{0.5}{mm}$ is used. The contact formulation needed is defined via asymmetric contact elements, neglecting surface effects like crack surface roughness. A rigid target plate (SOLID185, element size: $\SI{1}{mm}$) is defined for the MT-160 model, with frictionless rigid-to-flexible contact set between the MT-160 model and the rigid target surface. The crack surface uses CONTA174 overlaying the bottom surface of the MT-160 FE model and TARGE170 elements on top of the rigid target surface. The Augmented Lagrange Algorithm serves as the contact solver, with contact stiffness $\SI{1}{N/mm^2}$ and penetration limit $\SI{1}{\mu m}$ to minimize penetration. All initial gaps are closed. To analyse the impact of the near-plastic-zone integration, the FE model can include both linear-elastic and elastic-plastic material behavior using bilinear isotropic hardening, based on the properties of AA2024-T3. The material properties, used for the material model, are given in Table \ref{tab:material_properties}.
\begin{table}[h]
	\centering
	\caption{Material properties of AA2024-T3, taken from \cite{Tamarin.2002}}
	\label{tab:material_properties}
	\begin{tabular}{|l c c|}
		\hline
		Young's Modulus & $E$ & \SI{68 000}{MPa} \\
		Poisson's Ratio & $\nu$ & 0.33 \\
		Yield Strength & $R_{\mathrm{p,0.2}}$ & \SI{345}{MPa} \\
		Tangent Modulus & $m_{\mathrm{t}}$ & \SI{984}{MPa} \\
		\hline
	\end{tabular}
\end{table}
In order to separate the effects of PZ and plastic wake, the authors perform both simulations with and without crack propagation (see Figure \ref{fig:crack_growth_algorithm}). In the latter case, the crack length remains constant throughout the entire simulation, i.e., no crack growth occurs (see Figure \ref{fig:crack_growth_algorithm}a). This approach offers several advantages: both the primary PZ and the cyclic PZ can be resolved. This method bases on the assumption that crack growth is negligibly small. All simulations without crack propagation include a total of ten loading cycles, each consisting of a loading and unloading step. The maximum load and load ratio are set to $F_\mathrm{max}=\SI{15}{kN}$ and $R=0.1$ comparable to the fatigue crack growth experiments in Section \ref{sec:fcp_experiments} below. This study implements the Releasing-Constraint Method for conducting crack propagation within the FE simulation. For each loading cycle, an additional step is added where the boundary conditions of an element row ahead of the current crack front in the crack plane are released. Thus, the crack grows by one element length per cycle. The principle is schematically shown in Figure \ref{fig:crack_growth_algorithm}b. Between crack growth steps, the crack is subjected to only one loading and unloading sequence, despite literature \cite{Antunes.2008, Pommier.2000, Pommier.2001} recommending at least two cycles. This choice aims to save computation time as is not the focus of this paper to determine accurate crack opening SIFs ($K_\mathrm{op}$), supported by \auths{Camas} \cite{Camas.2019, Camas.2020}, who assessed the impact as being negligible. The crack propagation is performed within the refined element region. The boundary conditions are released at minimum load $F_{\mathrm{min}}$, referencing Laird's \cite{Laird.1977} observations of crack tip blunting at minimum load. A total of 40 loading cycles (150 steps) are simulated, resulting in a crack length of $a_\mathrm{Rifo}=\SI{2}{mm}$. That means that the starting crack length is $a-a_\mathrm{Rifo}$.

In order to evaluate the interaction integral in a comparable way to the evaluation in HR-DIC data, the nodal displacement data on the free surface of the FE simulation is extracted. In the finite element (FE) model, only the displacements above the crack are calculated. Consequently, the nodal displacements are symmetrically mirrored to the crack's underside, with the crack itself serving as the mirror axis. The procedure is schematically illustrated in Figure \ref{fig:fe_model}.

\subsection{Fatigue crack growth experiments}
\label{sec:fcp_experiments}
\begin{figure}[ht]
	\centering
	\includegraphics[width=0.95\textwidth]{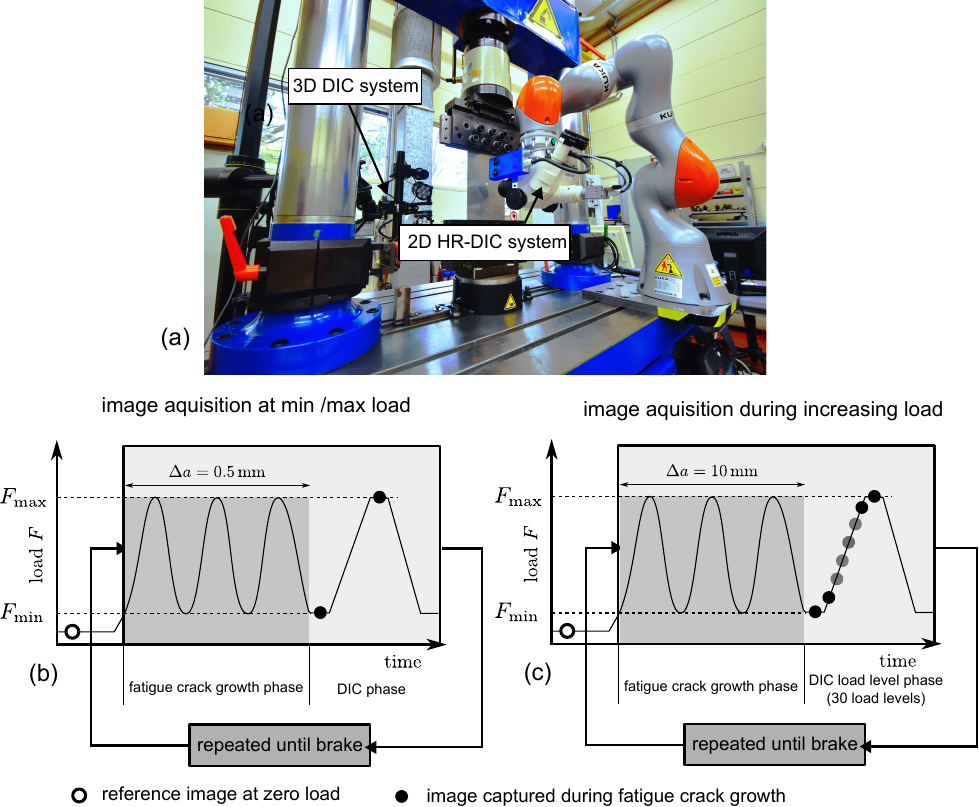}
	\caption{Fatigue crack growth experiments on MT-160 specimens of AA2024-T3 with multi-scale DIC measurements: (a) Experimental setup in the labotory, (b) measurement sequence every $\Delta a=\SI{0.5}{mm}$, (c) measurement sequence every  $\Delta a=\SI{10}{mm}$; at each image trigger point both backside DIC und the HR-DIC system are activated.}
	\label{fig:experimental_setup}
\end{figure}

A fatigue crack growth experiment was conducted as a comparison to the FE results on a 2-mm-thick MT-160 specimen produced from AA2024-T3. The fatigue crack propagates vertical the the rolling direction of the sheet material (L-T crack orientation). The SIF behaviour of a mode-I fatigue crack can be expressed by the following Formula that refers to ASTM E647-15 \cite{ASTM.E647}:
\begin{equation}
	\label{eq:mt_LEBM}
	K_\mathrm{norm}=\frac{F}{t} \sqrt{\frac{\pi \alpha}{2 W} \sec \left( \frac{\pi \alpha}{2}\right)}
\end{equation}
$\alpha$ denotes the ratio of crack length $a$ and the specimen width $W$. 
The experimental setup is illustrated in Figure \ref{fig:experimental_setup}a, utilizing a servo-hydraulic testing machine to apply sinusoidal loading with a constant amplitude. The load parameters are consistent with the numerical studies, with $F_\mathrm{max}=\SI{15}{kN}$ and $R=0.1$. The test stand is equipped with a robot-assisted HR-DIC system (Figure \ref{fig:experimental_setup}a) \cite{Paysan.2023}. This system features a KUKA iiwa robot that moves a Zeiss 206C stereo microscope, equipped with a Basler a2A5320-23umPro global shutter 16-megapixel CMOS camera, allowing for HR-DIC displacement field measurements around the crack tip during crack propagation. A detailed overview of the setup, including algorithms for optimizing image quality and reducing DIC noise, is provided in \cite{Paysan.2023}. 
Prior to the fatigue crack growth experiment, the robot moves the microscope to all predefined positions on the specimen surface where the crack is expected to propagate. The area is covered in a checkerboard pattern at zero load, and a reference image is captured at each position. Automatic algorithms ensure precise alignment with the specimen surface, as well as optimal sharpness and contrast of the images. After this calibration phase, the crack propagation phase begins. The specimen is cyclically loaded until a crack length increment of $\Delta a = \SI{0.5}{mm}$ is measured using a direct current potential drop (DCPD) system ($I=\SI{100}{mA}, U=\SI{60}{mV}$). The system then pauses, and the robot moves to the reference position that best captures the crack tip area. In the following an image at maximum and minimum load is taken (see Figure \ref{fig:experimental_setup}b). This sequence is repeated every $\Delta a = \SI{0.5}{mm}$. Once this sequence repeated 20 times, such that a crack length increment of $\Delta a = \SI{10}{mm}$ has been achieved, an entire loading step is captured by in total 30 load levels at which an image is captured. This is required in order to analyse the influence of crack closure on the results of the interaction integral. As with the previous sequence, this sequence is repeated until final failure of the MT-160 specimen. 
In addition to the robot-assisted HR-DIC system on the front side of the specimen, a 3D DIC system is located on the backside, enabling to capture displacement fields at the macroscopic level. Here, a GOM Aramis 12M 3D DIC System is used. The 3D DIC system consists of two individual 12-megapixel cameras with a distance of $\SI{98}{mm}$ and angles $\SI{25}{^\circ}$ relative to each other. The measurement volume is set to $\SI{200}{mm} \times \SI{150}{mm} \times \SI{21}{mm}$. 
The settings for the DIC evaluation and the corresponding resolutions of the 3D DIC and 2D HR-DIC systems are provided in Table \ref{tab:auflösung_dic}. The DIC evaluation is performed using the licensed Zeiss GOM Aramis 2020 software \cite{ARAMIS.2020}.

\begin{table}[htbp!]
	\centering
	\begin{tabular}{|lcc|}
		\hline
		Parameter & 3D DIC System & 2D HR-DIC System  \\ [0.5ex]
		\hline
		Measurement area  & $\SI{330}{mm}$ x $\SI{160}{mm}$ & $\SI{10.2}{mm}$ x $\SI{5.4}{mm}$ \\ 
		Facet spacing  & $\SI{16}{\text{pixels}}$ x $\SI{16}{\text{pixels}}$ & $\SI{30}{\text{pixels}}$ x $\SI{30}{\text{pixels}}$ \\ 
		Facet size  & $\SI{21}{\text{pixels}}$ x $\SI{21}{\text{pixels}}$ & $\SI{40}{\text{pixels}}$ x $\SI{40}{\text{pixels}}$ \\ 
		Spatial resolution  & $\SI{0.59}{mm/facet}$ & $\SI{47}{\mu m/facet}$ \\
		\hline
	\end{tabular}
	\caption{Resolution and facet settings for DIC evaluation of the 3D DIC and 2D HR-DIC measurement systems.}
	\label{tab:auflösung_dic}
\end{table}

%%%%%%%%%%%%%%%%%%%%%%%%%%%%%%%%%%%%%%%%%%%%%%%%%%%%%%%%%%%%%%%%%%%%%%%%%%%%%%
\section{Results and discussion} \label{sec:results}
%%%%%%%%%%%%%%%%%%%%%%%%%%%%%%%%%%%%%%%%%%%%%%%%%%%%%%%%%%%%%%%%%%%%%%%%%%%%%%
The objective of this study is to identify the characteristics that must be considered when applying the interaction integral to HR-DIC data. To ensure that the identified phenomena not only account for a single data point, the authors first investigate the determined SIF results along the entire fatigue crack. Subsequently, it is attempted to separate the individual systematic effects by comparing the HR-DIC result with numerical results at a crack length of $a=\SI{27.8}{mm}$ using this data point as a reference being exemplary for all.

\subsection{Fatigue crack growth curve and crack tip stresses}
In general, within this section,  two different kinds of crack propagation curves are distinguished: $da/dN-\Delta K$-curves according to ASTM E647-15 \cite{ASTM.E647} and $da/dN-\Delta K_{\mathrm{DIC}}$-curves that are constructed based on DIC and HR-DIC data. The following two paragraphs provide an overview over their construction procedures.

\textbf{Crack propagation curve according to ASTM E647-15}
\begin{adjustwidth}{1em}{}
The crack length is determined through potential drop measurements, and the crack growth rate is assessed using the secant method with crack growth intervals of $\Delta a = \SI{0.1}{mm}$. From the use of potential drop measurement follows that both crack tips (left and right in regards to the center notch) are taken into account for the crack length determination. A separate analysis of each side is not possible and the crack is assumed to grow straight. Finally, $K_\mathrm{norm}$ is computed according to Equation \ref{eq:mt_LEBM}.
\end{adjustwidth}

\begin{figure}[ht]
	\centering
	\includegraphics[width=0.6\textwidth]{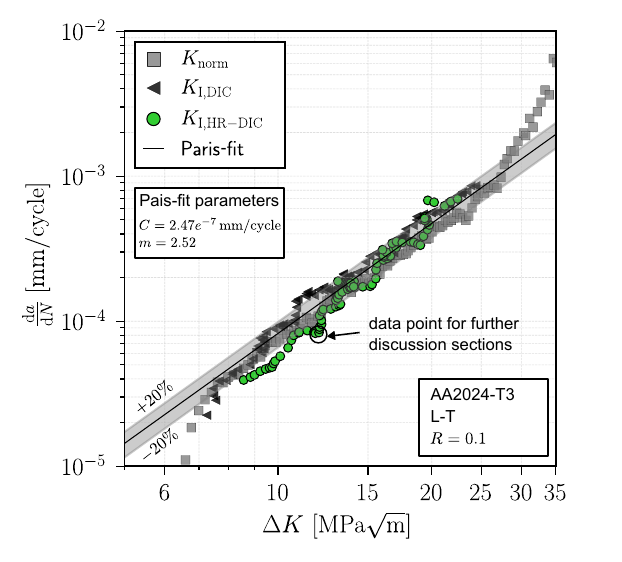}
	\caption{Fatigue crack growth curve $da/dN-\Delta K$ evaluated according to ASTM E647-15 \cite{ASTM.E647} and DIC und HR-DIC data; AA2024-T3 in L-T crack orientation at $R=0.1$.}
	\label{fig:fcp_curve}
\end{figure}

\textbf{DIC and HR-DIC crack propagation curves}
\begin{adjustwidth}{1em}{}
Figure \ref{fig:fcp_curve} presents the crack growth curves obtained from DIC and HR-DIC measurements. Images at maximum and minimum load levels were captured at intervals of $\SI{0.5}{mm}$. The displacement field data at both load levels were employed to calculate the crack tip SIFs, $K_\mathrm{I,DIC}$ and $K_\mathrm{I,HR-DIC}$, using the interaction integral method outlined in Section \ref{sec:interaction_integral}. The difference between these two values provides the cyclic SIF $\Delta K$. For the determination of $K_\mathrm{I,DIC}$, 20 integration paths are evaluated. The closest path to the crack tip maintains an upper, lower, and lateral distance of $\SI{5}{mm}$. The distance between subsequent paths is $\SI{1}{mm}$, excluding the crack path itself. An example of the integration paths in the macroscopic DIC data is given in Figure \ref{fig:dic_hrdic_comparison}a.
The definition of the integration paths in the HR-DIC data, as shown in Figure \ref{fig:dic_hrdic_comparison}b, is not fixed. Instead it is adjusted according to the dimensions of the PZ. The PZ is estimated by applying the yield stress as a threshold. An additional offset of $\SI{0.5}{mm}$ is applied, setting the minimum distance of the first integration path from the crack tip. The spacing between paths is $\SI{0.2}{mm}$. The  gap around the crack path in the integration paths is modified to account for the dimensions of the plastic wake. The median of the individual path results determines the final $K_\mathrm{I,DIC}$ or $K_\mathrm{I,HR-DIC}$ results. Crack growth rate measurements are based on crack length data obtained using the crack tip detection model ParallelNets \cite{Melching.2022}, which is part of the CrackPy package \cite{CrackPy.2022}. Therefore, the crack lengths for both the DIC and HR-DIC curves are derived from the displacement field data. Crack lengths are smoothed using a moving average over three consecutive points, and the slope between three consecutive data points is calculated to determine the crack growth rate $da/dN$.
\end{adjustwidth}

Figure \ref{fig:fcp_curve} shows the crack growth curves based on DIC and HR-DIC measurements as well as the FCP curve according to ASTM E647-15 \cite{ASTM.E647}.

From a methodology point of view, both different FCP curve types are difficult to compare since the FCP curves according to ASTM E647-15 \cite{ASTM.E647} incorporates the growth rate of two crack tips, while the DIC and HR-DIC based ones only consider the crack tip of the right side of the MT-160 specimen. Nevertheless, from a qualitative perspective, all curves shown in Figure \ref{fig:fcp_curve} reveal the linear behavior on a double-logarithmic scale. This linearity confirms the applicability of the Paris law for describing the crack growth behavior in all presented curves, as most sections of all curves fall within a common 20\% scatter band that is often used as reproducible criterion for $da/dN-\Delta K$-curves of the same material. This shows the redundancy of the methods to each other. In order to assess the deviation of the curves in the Paris regime, all available data points (ASTM, DIC and HR-DIC) are used within the interval $8 \, \mathrm{MPa\sqrt{m}} < \Delta K < 25\, \mathrm{MPa\sqrt{m}}$ for fitting the Paris law. The fit is illustrated in Figure \ref{fig:fcp_curve} and is defined by $C=2.47e^{-7}$ and $m=2.52$ as fitting parameters, resulting in the formula
\begin{equation} \label{eq:paris_fit}
    \frac{da}{dN} = C \left( \Delta K \right)^m \quad \bigg[ \Leftrightarrow \quad \log \left( \frac{da}{dN} \right) = \log(C) + m \cdot \log(\Delta K) \bigg].
\end{equation}
Those fitting parameters are consistent to literature values \cite{Bergner.2001} for the L-T crack orientation of AA2024-T3. In the following, each curve is evaluated for its deviation against Equation \eqref{eq:paris_fit}. The results are given in Table \ref{tab:fcp_curve_deviation}.

\begin{table}[h]
    \centering
    \begin{tabular}{|l c c|}
        \hline
        curve type & $R^{2}$ [-] & Proportion outside the 20\,\% scatter band [\%] \\ \hline
        ASTM curve & 0.99 & 3\,\% \\ 
        DIC curve & 0.98 &  5\,\%\\ 
        HR-DIC curve & 0.96 & 17\,\% \\ 
        \hline
    \end{tabular}
    \caption{Deviation analysis of the single FCP curves; determined against  a Paris fit with $C=2.47e^{-7}$ and $m=2.52$ and evaluated within the SIF interval of $\Delta K=[8,\, 25]\, \mathrm{MPa\sqrt{m}}$}
    \label{tab:fcp_curve_deviation}
\end{table}

\begin{figure}[ht]
	\centering
	\includegraphics[width=0.95\textwidth]{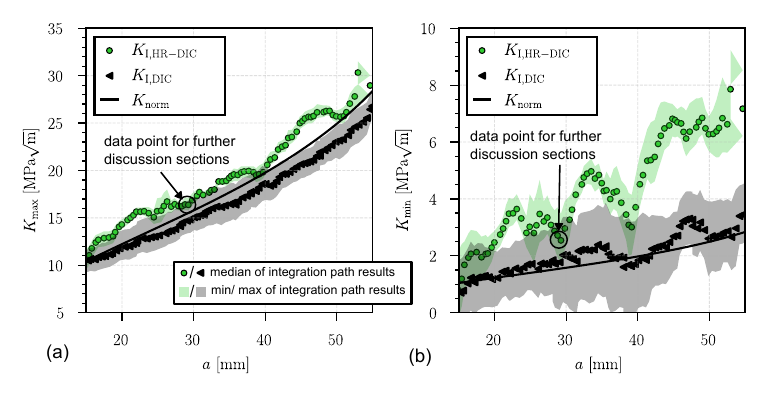}
	\caption{Evolution of the maximum SIF $K_\mathrm{max}$ and minimum SIF $K_\mathrm{min}$ along the crack length compared to the analytical SIF solution according to ASTM E647-15 \cite{ASTM.E647} (see Equation \ref{eq:mt_LEBM})}
	\label{fig:sif_experiment}
\end{figure}

Generally, all fatigue crack propagation (FCP) curves show a good agreement with each other as quantified by the large $R^{2}$-values. While only a few data points of the ASTM and DIC FCP curves are located outside the 20\,\% scatter band, this value is increased in case of the HR-DIC FCP curve. Comparing both the DIC and HR-DIC FCP curves, it can be recognised that the HR-DIC one always slightly shifted to the right, indicating that at larger cyclic crack tip stress lower growth rates would exist. Since the growth rates in DIC und HR-DIC are almost identical in double-logarithmic scale, this phenomenon mainly results from different measured SIF values as shown in Figure \ref{fig:sif_experiment}. 

Figure \ref{fig:sif_experiment} illustrates the variation of the measured $K_\mathrm{I,DIC}$ and $K_\mathrm{I,HR-DIC}$ results at maximum load $F_\mathrm{max}$ and minimum load $F_\mathrm{min}$ along the crack length $a$. The scatter bands in both graphs represent the maximum and minimum $K_\mathrm{I}$ values from the integration path results, indicating that the application of the interaction integral based on HR-DIC data results in a reduction in scatter of the path results compared to those derived from DIC data. Consequently, HR-DIC data not only enables the identification of microscopic effects in fatigue cracks but also its crack tip stress determination is less affected by the DIC measurement noise.
Furthermore, Figure \ref{fig:sif_experiment} provides an explanation for the shift of HR-DIC FCP curves to the right compared to the DIC-based FCP curves. Figure \ref{fig:sif_experiment}a illustrates that $K_\mathrm{I,HR-DIC}$ at maximum load exceeds $K_\mathrm{I,DIC}$ by 2 to $6 \, \mathrm{MPa\sqrt{m}}$. This results in a shift of the HR-DIC-based crack growth curves below the two remaining crack growth curves, particularly in the range $\Delta K \in [5,\, 15] \, \mathrm{MPa\sqrt{m}}$. The subsequent sections analyze the reasons for the deviation of $K_\mathrm{I,HR-DIC}$ compared to the other two curve trends and provide recommendations for the path definition for the application of the interaction integral in HR-DIC displacement field data.

\subsection{The influence of integrating near the PZ}
\label{sec:sif_plastic_zone}
\begin{figure}[ht]
	\centering
	\includegraphics[width=0.95\textwidth]{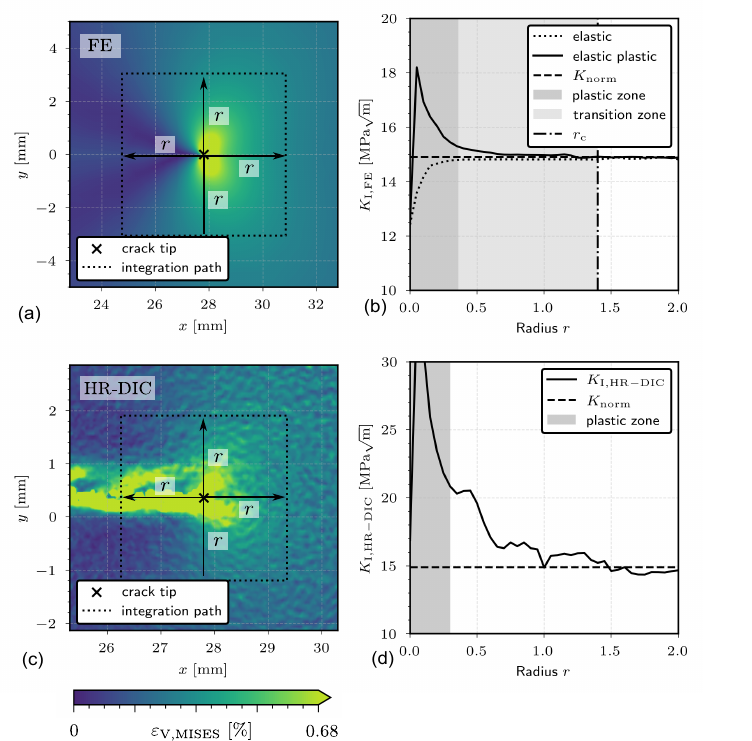}
	\caption{Influence study analyzing the impact of integration in close proximity of the PZ: (a) FE von MISES equivalent strain on the free surface of the elastic-plastic FE model, (b) $K_\mathrm{I,FE}$ path results for different $r$, (c) HR-DIC von MISES equivalent strain and (d) $K_\mathrm{I,HR-DIC}$ path results for different $r$; evaluated at $K_\mathrm{norm} = \SI{14.9}{MPa\sqrt{m}}$ and $a=\SI{28}{mm}$.}
	\label{fig:sif_plastic_zone}
\end{figure}

Due to the limited measurement range of HR-DIC systems, the integration paths for the interaction integral are in close proximity to the PZ. Since plastic deformation affects the crack tip field near the PZ, the behaviour deviates from pure linear-elastic conditions. In order to investigate the impact of integrating in close proximity to the PZ, the authors employ the FE model from Section \ref{sec:fe_simulation}. Here, no crack propagation is conducted during the load cycles. In regards to the material model implemented, both linear-elastic (elastic modulus $E=73.1$\,GPa, Poisson ratio $\nu=0.33$) and a elastic-plastic material model are used separately. In the latter case, a bilinear isotropic hardening material model (yield stress $\sigma_\mathrm{yield}=\SI{350}{MPa}$, hardening modulus  $m_\mathrm{t}=\SI{984}{MPa}$) is employed. To evaluate the influence of the PZ on $K_\mathrm{I}$, a total number of 200 quadratic closed integration paths are analyzed, starting at a distance of $r=\SI{0.05}{mm}$ from the crack tip position. For each integration path $r$ is increased by $\Delta r=\SI{0.05}{mm}$. Consequently, the last integration path is located at $r=\SI{10}{mm}$. Figure \ref{fig:sif_plastic_zone} compares the $K_\mathrm{I}$ path results based on FE data $K_\mathrm{I,FE}$ coming from elastic or elastic-plastic simulations with those results derived from HR-DIC displacement field data $K_\mathrm{I,HR-DIC}$.

\begin{figure}[ht]
	\centering
	\includegraphics[width=0.95\textwidth]{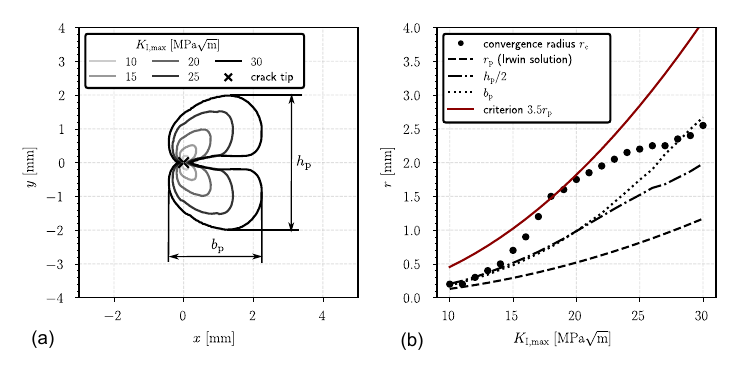}
	\caption{The evolution of the convergence radius $r_\mathrm{c}$ in dependency of different maximum SIFs: (a) the dimension and shape of PZ at selected maximum SIFs and (b) height $h_\mathbf{p}$ and width $b_\mathrm{p}$ of the PZ compared to the $r_\mathrm{c}$}
	\label{fig:convergence_study}
\end{figure}

Figures \ref{fig:sif_plastic_zone}b and d illustrate the convergence behaviour of the path results based on FE and HR-DIC displacement field data for $K_\mathrm{norm} = \SI{14.9}{MPa\sqrt{m}}$ as a function of the radius $r$. The von MISES equivalent strain fields are depicted in Figures \ref{fig:sif_plastic_zone}a and c. Within the PZ (PZ), the assumption of Hooke's law in the context of the interaction integral leads to an overestimation of $K_\mathrm{I}$. The size of the PZ in both Figures \ref{fig:sif_plastic_zone}b and d is estimated using Irwins formula under plane stress condition $r_\mathrm{p} = \SI{0.36}{mm}$. This overestimation arises from the assumption of linear-elastic behavior in the derivation of the interaction integral. As a result of the inaccurate description of the material behaviour, the integral becomes path-dependent, as also emphasized by \auths{Xu} \cite{Xu.2022, Xu.2022b}. Outside the PZ, in close proximity of the PZ, the material exhibits purely linear-elastic behavior, enabling the application of the interaction integral.
However, both convergence analyses shown in Figures \ref{fig:sif_plastic_zone}b and d reveal that the path results based on the elastic-plastic FE simulation $K_\mathrm{I,FE}$ and HR-DIC data $K_\mathrm{I,HR-DIC}$ do not converge immediately, even when the material behavior should be adequately represented by Hooke's law. A transition region exists in which $K_\mathrm{I}$ is still larger than $K_\mathrm{norm}$. This is attributed to deviations in the crack tip field due to the presence of the PZ, compared to the linear-elastic solution \cite{Carka.2011}. As the radius $r$ increases, the influence of the PZ on the crack tip field diminishes, and the path results begin to converge. At a radius of $r = \SI{1.4}{mm}$, the influence of the PZ on the crack tip field becomes negligible. This specific radius is subsequently referred to as the convergence radius $r_\mathrm{c}$. Due to the numerical scatter, the path results also fluctuate slightly in the converged region ($r>r_\mathrm{c}$). The maximum path value of the last 50 integration path defines the boundary value. As shown in Figure \ref{fig:sif_plastic_zone}b the SIF results are overestimated in $r<r_\mathrm{c}$. Coming from the left ($r=0$\,mm) the first SIF results at radius $r$ that intersects this boundary value, is considered as $r_\mathrm{c}$.
Figure \ref{fig:convergence_study} investigates the dependency of $r_\mathrm{c}$ on different maximum SIFs. Therefore, in total 21 FE simulation with a maximum SIF range of $K_\mathrm{max}=[10,\, 30]\, \mathrm{MPa\sqrt{m}}$ were performed. At the end of each simulation, $r_\mathrm{c}$ is determined based on the nodal solution of the free surface of the 3D FE model. 

Figure \ref{fig:convergence_study} shows that $r_\mathrm{c}$ depends on the maximum SIF since it increases with increasing stress intensity. Furthermore, the results indicate that $r_\mathrm{c}$ is influenced by the shape of the PZ. From literature \cite{Besel.2016, Caputo.2013, Gadallah.2021}, it is known, that in thin sheet materials the wing shape of the PZ changes with increasing maximum SIF. This is referred to the increasing pre-dominant plane stress condition within the sheet material. At a specific ratio between the thickness of the sheet and the maximum SIF, the wings of the PZ expand more in width than in height as exemplary shown in Figure \ref{fig:convergence_study}a for the configuration considered in this study. In case of this analysis, this turning point is reached at $K_\mathrm{I,max}=\SI{17}{MPa\sqrt{m}}$ as shown in Figure \ref{fig:convergence_study}b. The shapes of the PZ at selected maximum SIF are given in Figure \ref{fig:convergence_study}a. Regarding $r_\mathrm{c}$, the values exhibit a quadratic increasing behavior until this turning point is reached. Afterwards, it increases almost parallel to the wing height of a single wing $h_\mathbf{p} / 2$.

It is therefore challenging to derive a general convergence criterion, given the numerous influences on the shape of PZ, including mixed-mode loading conditions, material and microstructure properties, and specimen thickness. To provide a conservative but robust estimate for use with the interaction integral on displacement field data of AA2024-T3, the authors have developed the following criterion:
% @David: Kann man das ggf. etwas besser formulieren?
% Passt denke ich. Das nach dem Gleichheitszeichen kommt von Irwin?
\begin{equation}
	\label{eq:convergence}
	r \geq 3.5 \cdot r_\mathrm{p} = \frac{1.75}{\pi} \cdot \left( \frac{K_\mathrm{I,max}}{\sigma_\mathrm{yield}} \right)^{2}
\end{equation}
With satisfying this criterion, the influence of PZ on the SIF results can be neglected.
However, due to the limited measurement range of the HR-DIC data, a significant portion of the integration paths lies within the range $r_\mathrm{p,ESZ} < r < r_\mathrm{c}$ and, thus, $K_\mathrm{I,HR-DIC}$ is overestimated. This effect is particularly evident in the $K_\mathrm{max}$ over $a$ shown in Figure \ref{fig:sif_experiment}a. There is a systematic offset between $K_\mathrm{I,HR-DIC}$ and $K_\mathrm{norm}$ which increases slightly with increasing $K_\mathrm{max}$. As the size of the PZ increases, the proportion of integration paths within the convergence range decreases, explaining this increased overestimation. Furthermore, comparing the distances between $K_\mathrm{I,HR-DIC}$ and $K_\mathrm{norm}$ at maximum load (see Figure \ref{fig:sif_experiment}a) and minimum load (see Figure \ref{fig:sif_experiment}b) indicates that a load dependency exists. When the load is at its maximum, the elastic field around the crack tip is subject to a stronger influence, which in turn results in a more pronounced overestimation of $K_{\mathrm{I,HR-DIC}}$ at the maximum load in comparison to the minimum load. This effect may be the reason why the HR-DIC FCP curve in Figure \ref{fig:fcp_curve} is slightly shifted to the right compared to the DIC curve. 

\subsection{The impact of plastic wake and the integration path gap}
\begin{figure}[ht]
	\centering
	\includegraphics[width=0.95\textwidth]{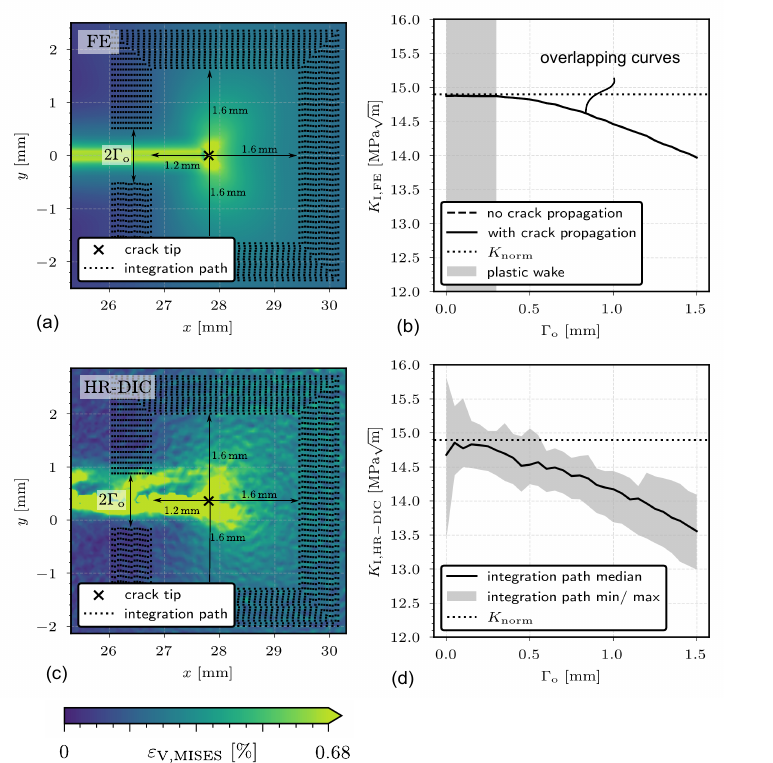}
	\caption{Influence study analysing the impact of the integration path gap and integrating within the plastic wake: (a) FE von MISES equivalent strain field of the free surface of the FE model with crack propagation, (b) $K_\mathrm{I,FE}$ path results for different integration path gaps $\Gamma_\mathrm{0}$, (c) HR-DIC von MISES equivalent strain field and (d) $K_\mathrm{I,HR-DIC}$ path results for different integration path gaps $\Gamma_\mathrm{0}$; evaluated at $K_\mathrm{norm} = \SI{14.9}{MPa\sqrt{m}}$ and $a=\SI{28}{mm}$}
	\label{fig:sif_plastic_wake}
\end{figure}

When applied to DIC displacement field data, closed line integrals are not feasible as the separation of material along the crack path results in erroneous displacement and strain calculations. These errors would distort the outcome of the interaction integrals. Furthermore, the HR-DIC system allows to resolve the plastic wake. Due to the locally varying crack path resulting from microscopic crack deflections and branching, it is often unavoidable, even with the integration path gap of $\Gamma_\mathrm{o}$, that portions of the integration path extend into the plastically deformed region around the crack path. At these integration points, the stresses $\sigma$ are therefore inaccurately calculated due to the underlying linear-elastic material model. In order to analyse the impact of the integration path gap as well as a likely integration within the plastic wake, FE simulations with and without crack propagation were conducted. Therefore, the Releasing-Constraint method from Section \ref{sec:fe_simulation} is used. The elastic-plastic material model is identical to the one in Section \ref{sec:sif_plastic_zone}. For integration, 10 integration paths are used with a distance of $\SI{0.05}{mm}$ to each other. The distances to each side of the rectangular integration path within the free surface von MISES equivalent strain field based on FE and HR-DIC are given in Figure \ref{fig:sif_plastic_wake}a and b. Regarding $K_\mathrm{norm}=\SI{14.9}{MPa\sqrt{m}}$, all integration paths satisfy the convergence criterion from Equation \ref{eq:convergence}.

Figure \ref{fig:sif_plastic_wake}b shows the variation of $K_\mathrm{I,FE}$ as a function of $\Gamma_\mathrm{o}$. Both trends, with and without crack propagation calculation, align precisely from a qualitative perspective. This observation indicates that the result of $K_\mathrm{I,FE}$ is not influenced by plastic wake. Instead, the study reveals sensitivity to the path gap $\Gamma_\mathrm{o}$. For $\Gamma_\mathrm{o} < \SI{0.5}{mm}$, $K_\mathrm{I}$ deviates by approximately $\SI{4}{\%}$ from $K_\mathrm{norm}$. The gradient of this decrease increases with the size of $\Gamma_\mathrm{o}$.
The comparison with the HR-DIC-based $K_\mathrm{I,HR-DIC}$ evaluation exhibits a nearly identical characteristic. Due to the excellent agreement between the trends of the FE and HR-DIC analyses, this study recommends keeping $\Gamma_\mathrm{o}$ as small as possible. To ensure the applicability of Hooke's law as a material model, it is strongly advised to define the integration paths as close as possible to the plastic wake, but within the elastic near-field. The analysis shows that even partial integration within the plastic wake results in a negligible deviation from $K_\mathrm{norm}$, despite being theoretically inaccurate from a continuum mechanics perspective. The analysis of the scatter band indicates that variations in the integration path results significantly increase when integrating close to the crack path ($\Gamma_\mathrm{o} = \SI{0}{mm}$). For the application of the interaction integral with the given HR-DIC data, a minimum gap of $\Gamma_\mathrm{o} > \SI{0.2}{mm}$ is recommended. As $K_\mathrm{max}$ increases, the size of the integration path gap $\Gamma_\mathrm{o}$ also increases. Consequently, the proportion of the gap relative to the entire integration path increases, leading, as shown in Figures \ref{fig:sif_plastic_wake}b and d, to lower $K_\mathrm{I}$ values than those calculated by \eqref{eq:mt_LEBM}. This effect partially compensates for the overestimation of $K_\mathrm{I}$ due to the influence of the PZ and its transition zone at high $K_\mathrm{max}$, resulting in HR-DIC-based FCP curves  aligning closely with $K_\mathrm{norm}$ in the range $\Delta K > \SI{15}{MPa\sqrt{m}}$ in Figure \ref{fig:fcp_curve}.

\subsection{The crack closure effect}
\label{sec:sif_crack_closure}
\begin{figure}[ht]
	\centering
	\includegraphics[width=0.95\textwidth]{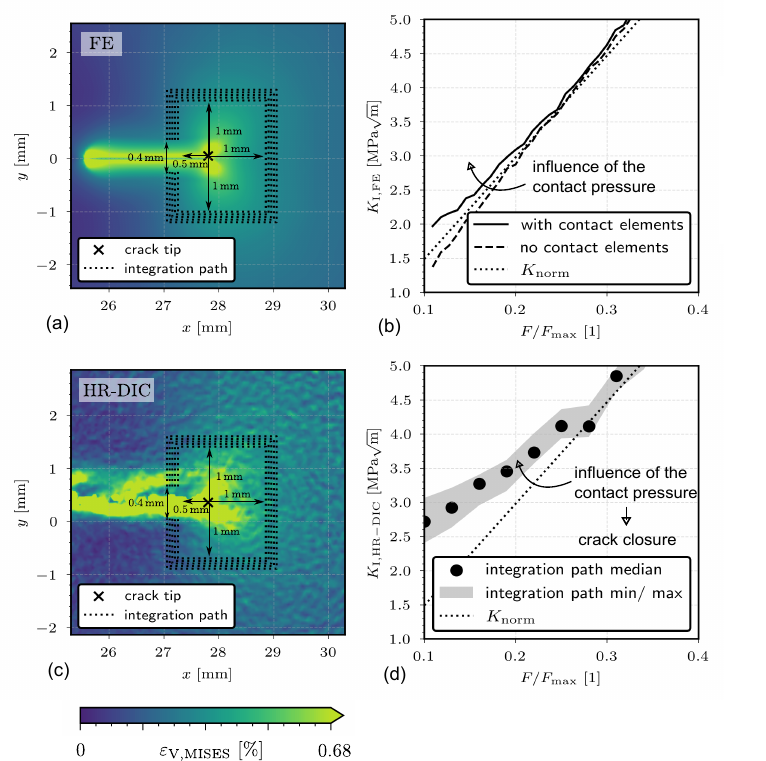}
	\caption{Influence study analysing the impact of crack face contact due to crack closure: (a) FE von MISES equivalent strain field of the free surface of the FE model, (b) $K_\mathrm{I,FE}$ as a function of the load $F$ comparing a FE simulation with and without contact definition, (c) HR-DIC von MISES equivalent strain field and (d) $K_\mathrm{I,HR-DIC}$ along the crack opening load $F$; evaluated at $K_\mathrm{norm} = \SI{14.9}{MPa\sqrt{m}}$ and $a=\SI{28}{mm}$}
	\label{fig:sif_crack_closure}
\end{figure}

The original formulation of the interaction integral assumes crack flanks to be free of external loads. However, this assumption is no longer valid in the presence of crack closure. \auth{Elber} \cite{Elber.1970b} demonstrated that, during fatigue crack growth in AA2024-T3 at $R=0.1$, plasticity-induced crack closure occurs. Moreover, several numerical studies \cite{Antunes.2019, Masuda.2021, Oplt.2023} have shown that contact between the crack flanks primarily occurs at the free specimen surface. The contact pressure induces a residual stress field, which affects the application of the interaction integral at the surface.
The question arises as to how crack face contact, resulting from crack closure mechanisms, influences the outcome of $K_\mathrm{I}$. To address this, the elastic-plastic FE crack propagation simulation from the previous study is extended to include the contact definition by using contact elements. The evaluation of $K_\mathrm{I,FE}$ is based on the nodal displacements of the specimen surface. To isolate the effect of crack face contact from other influences discussed in previous sections, $K_\mathrm{I,FE}$ is determined both with and without the contact definition. For the simulation without contact definition the rigid target surface as well as all contact elements are deleted, allowing the crack faces of the MT-160 model to deform freely. This approach allows for a targeted investigation of the influence of crack face contact and the resulting elastic residual stress field near the crack path. The results are compared with the HR-DIC-based evaluation of $K_\mathrm{I,HR-DIC}$. For this purpose, $K_\mathrm{I,HR-DIC}$ is determined at each of the 30 load steps during crack opening. The HR-DIC data are derived from the load cycle recording at $a = \SI{27.8}{mm}$. The location of the ten integration paths is identical in both the FE and HR-DIC data, as shown in Figures \ref{fig:sif_crack_closure}a and c. 

\begin{table}[h]
    \centering
    \begin{tabular}{|c c c|}
        \hline
        integration path & $K_\mathrm{I,FE}$ without contact & $K_\mathrm{I,FE}$ with contact \\ \hline
        \#1 & $\SI{0.69}{MPa\sqrt{m}}$ & $\SI{1.53}{MPa\sqrt{m}}$\\ 
        \#2 & $\SI{1.14}{MPa\sqrt{m}}$ &  $\SI{1.81}{MPa\sqrt{m}}$\\ 
        \#3 & $\SI{1.37}{MPa\sqrt{m}}$ &  $\SI{1.97}{MPa\sqrt{m}}$\\ 
        \#4 & $\SI{1.60}{MPa\sqrt{m}}$ &  $\SI{2.16}{MPa\sqrt{m}}$\\ 
        \#5 & $\SI{1.82}{MPa\sqrt{m}}$ &  $\SI{2.36}{MPa\sqrt{m}}$\\ 
        \hline
    \end{tabular}
    \caption{$K_\mathrm{I,FE}$ results of the single integration paths evaluated on displacement field data of the free surface of the 3d FE model at minimum load $K_\mathrm{norm}=\SI{1.49}{MPa\sqrt{m}}$; integration path \#1 denotes the path that is closest to the crack tip}
    \label{tab:sif_crack_closure_path_results}
\end{table}

%%%%%%%%%%%%%%%%%%%%% !!!!!!!!
% Ich bin von den verschiedenen K's verwirrt. K_cp und K_op ist nicht definiert.
% Ausserdem müssen wir aufpassen, dass immer klar ist was wir bei K_FE meinen - mit oder ohne Kontakt?
%%%%%%%%%%%%%%%%%%%%%
%%%%%%%%%%%%%%%%%%%%%
Figure \ref{fig:sif_crack_closure}b compares the $K_\mathrm{I,FE}$ trends based on FE data with and without the contact definition. The results indicate that crack face contact leads to an increase in $K_\mathrm{I,FE}$ compared to $K_\mathrm{norm}$. This behavior is not observed in the $K_\mathrm{I,FE}$ trend without the contact definition, suggesting that crack face contact results in an increased $K_\mathrm{I,FE}$. The comparison with the HR-DIC-based evaluation of $K_\mathrm{I,HR-DIC}$ shows a similar trend, consistent with the findings of $J$-integral studies by \auths{Gonzales} \cite{Gonzales.2017}. The main difference is that the deviation from the $K_\mathrm{norm}$ reference line is more pronounced. This is also evident in the comparison of $K_\mathrm{I}$ values at minimum load $F_\mathrm{min}$. Specifically, $K_\mathrm{I,FE,min}=\SI{2}{MPa\sqrt{m}}$ is lower than $K_\mathrm{I,HR-DIC,min}=\SI{2.74}{MPa\sqrt{m}}$. This difference is attributed to the significantly higher contact pressure between the two crack faces. Compared to the FE crack propagation simulation, the actual crack growth increment per load cycle is much smaller than $\Delta a = \SI{0.08}{mm}$. Consequently, the accumulation of plastic strain within the PZ is substantially increased for propagating fatigue cracks in AA2024-T3, which also results in larger contact pressures between the crack faces. Furthermore, the good agreement between the FE-based analysis and the HR-DIC data-based evaluation suggests that the predominant crack closure mechanism is plasticity-induced crack closure. This conclusion is based on the fact that the FE model is only capable of representing this specific closure mechanism.
Furthermore, Table \ref{tab:sif_crack_closure_path_results} indicates that the crack closure contact does not lead to an increase in path dependency. The difference values between the successive single-path results are almost all of a similar order of magnitude.

The authors investigated the crack closure behavior of the identical crack tip and specimen in a previous study using local COD measurements to analyse the crack opening kinematics \cite{Paysan.2024}. Therefore, the $K_\mathrm{op}$ values along the crack front for the same numerical model presented in Section \ref{sec:fe_simulation} and the same HR-DIC measurements were investigated. $K_\mathrm{op,ctod}$ denotes the first crack tip stress at which no contact between the crack faces close behind the crack tip exists during crack opening. Table \ref{tab:kop_results} presents the $K_\mathrm{op,ctod}$ values for the FE and HR-DIC investigation. 
\begin{table}[h]
    \centering
    \begin{tabular}{|l c c|}
        \hline
        SIF & FE & HR-DIC \\ \hline
        $K_\mathrm{op,ctod}$ $\mathrm{MPa\sqrt{m}}$ & 6.63 & 7.42\\
        $K_\mathrm{I,min}$ $\mathrm{MPa\sqrt{m}}$ & 1.97 & 2.66\\
        \hline
    \end{tabular}
    \caption{Comparison between $K_\mathrm{I,min}$ from FE and HR-DIC displacement field data and the actual crack opening loads $K_\mathrm{op,ctod}$ (see \cite{Paysan.2024}); $K_\mathrm{I,min}$ represents the median of the path results}
    \label{tab:kop_results}
\end{table}
Table \ref{tab:kop_results} points out that the actual crack opening loads are significant higher, indicating that $K_\mathrm{I,min}$ derived by the interaction integral from displacement field data does not represent the crack opening load $K_\mathrm{op,ctod}$. However, the previous results showed that crack closure increases slightly the SIF value which causes additionally the increasing character of the $K_\mathrm{I,HR-DIC}$ trend along the crack length at minimum load from Figure \ref{fig:sif_experiment}b.

%%%%%%%%%%%%%%%%%%%%%%%%%%%%%%%%%%%%%%%%%%%%%%%%%%%%%%%%%%%%%%%%%%%%%%%%%%%%%%
\section{Conclusions}
%%%%%%%%%%%%%%%%%%%%%%%%%%%%%%%%%%%%%%%%%%%%%%%%%%%%%%%%%%%%%%%%%%%%%%%%%%%%%%

In this paper, the authors investigate the applicability of the interaction integral for determining SIFs within high-resolution displacement field data. Given the aim of the fracture mechanics community to obtain multi-scale time-series data that bridges the gap between local microscopic fatigue crack growth mechanisms and their impact on the macroscopic fatigue crack growth behaviour represented by the FCP curve, this paper identifies influences and analyses their impact on the interaction integral. In particular, the following conclusions are drawn:

\begin{itemize}
    \item HR-DIC and DIC-based crack propagation curves, developed using the new methodology, accurately reflect the crack growth behavior similar to the established ASTM E647-15 \cite{ASTM.E647} method. This is demonstrated by their agreement within a common scatter band and is quantitatively supported by high $R^{2}>0.95$ values, as shown in the comparison of the respective crack growth curves.
    
    \item Integrating near the PZ results likely in an overestimation of the SIF. Furthermore, a dependency of the convergence radius on the shape of the PZ is identified. The convergence radius describes the minimum distance between the integration path and the crack tip, and for consistent results, a conservative estimate can be obtained by using the condition $r \geq 7/2 \cdot r_\mathrm{p}$, where $r_\mathrm{p}$ is the PZ size.
    
    \item The size of the path gap around the crack path relative to the total integration path length significantly impacts the interaction integral results. The results indicate that larger gaps lead to smaller values. It is recommended to keep the exclusion gap as small as possible. Integrating close to the crack path increases the scatter between different integration paths pointing towards a loss in path independence. However, this seems to have a negligible effect on the mean result, provided the integration paths remain within the elastic field near the PZ.
    
    \item Crack face contact seems to increase the interaction integral results, but the values remain significantly lower than the measured $K_\mathrm{op}$ values. Moreover, the results indicate that crack face does not contribute to an increased loss of path independence.
\end{itemize}

%%%%%%%%%%%%%%%%%%%%%%%%%%%%%%%%%%%%%%%%%%%%%%%%%%%%%%%%%%%%%%%%%%%%%%%%%%%%%%
\section{Acknowledgements}
%%%%%%%%%%%%%%%%%%%%%%%%%%%%%%%%%%%%%%%%%%%%%%%%%%%%%%%%%%%%%%%%%%%%%%%%%%%%%%

The authors acknowledge the financial support of the DLR-Directorate Aeronautics. This work was supported by the Deutsche Forschungsgemeinschaft, Germany (DFG) via the project Experimental analysis and phase-field modeling of the interaction between PZ and fatigue crack growth in ductile materials under complex loading (grant number BR 6259/2-1). Furthermore, funding came from the Federal Ministry for Economic Affairs and Climate Action, Germany on the basis of a decision by the German Bundestag, within the framework of the aerospace program LuFo-VI of the project "Intelligent FSW Process Monitoring" (Funding ID 20W2201E).

%%%%%%%%%%%%%%%%%%%%%%%%%%%%%%%%%%%%%%%%%%%%%%%%%%%%%%%%%%%%%%%%%%%%%%%%%%%%%%
\section{Data availability}
Code and data will be published if the paper is accepted.

\section{Competing interests}
The authors declare no competing financial or non-financial interests.

\section{Author contributions}
F.P. conceived the idea, conducted the simulations and experiments, evaluated the results, joined the discussions and wrote the manuscript. D.M. joined the discussions and wrote the manuscript. E.B. joined the discussions and wrote the manuscript.

\printbibliography

\end{document}